# Laser Ablation and Injection Moulding as Techniques for Producing Micro Channels Compatible with Small Angle X-Ray Scattering

R. Haider[1], B. Marmiroli[1], I. Gavalas[2], M Wolf[1], M. Matteucci[3], R. Taboryski[3], A. Boisen[3], E. Stratakis[2] and H. Amenitsch[1]

[1]*Graz University of Technology, Institute of Inorganic Chemistry, Stremayrgasse 9/IV, 8010 Graz, Austria*
[2]*Institute of Electronic Structure and Lasers, Foundation for Research and Technology and University Of Crete, Department of Materials Science and Technology 100 Nikolaou Plastira Str. Vassilika Vouton Heraklion Crete GR-700 13, Greece*
[3]*DTU Nanotech, Technical University of Denmark , Department of Micro- and Nanotechnology, Ørsteds Plads, Building 345C, Building 423, 2800  Kgs. Lyngby, Denmark*

e-mail: amenitsch@tugraz.at, benedetta.marmiroli@tugraz.at

Microfluidic mixing is an important means for *in-situ* sample preparation and handling while Small Angle X-Ray Scattering (SAXS) is a proven tool for characterising (macro-)molecular structures. In combination those two techniques enable investigations of fast reactions with high time resolution (<1 ms).

The goal of combining a micro mixer with SAXS, however, puts constraints on the materials and production methods used in the device fabrication. The measurement channel of the mixer needs good x-ray transparency and a low scattering background. While both depend on the material used, the requirement for low scattering especially limits the techniques suitable for producing the mixer, as the fabrication process can induce molecular orientations and stresses that can adversely influence the scattering signal.

Not only is it important to find a production method that results in a device with low background scattering, but it also has to be versatile enough to produce appropriate mixer designs.

Here we discuss two methods – laser ablation of polycarbonate and injection moulding of Topas – which were found suitable for our needs, provided care is taken in aligning the mixing/reaction channel, where the actual measurements will be carried out. We find injection moulding to be the better of the two methods.

# 1 Introduction

Microfluidic mixers combine fast mixing times, high time resolution and low sample consumption. Small Angle X-Ray Scattering (SAXS) probes structures in the 1-100 nm range, providing e.g. conformal and structural information of nanoparticles, biological molecules etc. [1]. Combining those two gives a potent setting for time resolved investigations of fast reactions, like protein folding [2].

Our mixer under design should allow mixing of two reactants in 5 ms with a similar subsequent time resolution, probing up to a total reaction time of about 1 s, while using a few µl/s of total flow and a sample usage below the µl/s range. However, the reactants, especially organic ones, are often problematic as they tend to stick to and aggregate at the channel walls, causing data spoilage or even channel clogging. We have previously addressed those difficulties by using a free-jet mixer [3], but this device suffers from rapid jet break-up and consequently low signal-to-noise ratio. To circumvent this problem we are currently following an in-channel, 3D-sheathing approach. This is not entirely new (see e.g. [4]), but the intended application of the device in SAXS experiments causes certain difficulties.

On one hand the mixing and measurement channel should be narrow enough to facilitate fast mixing, high time resolution as well as low sample consumption. On the other hand it should be deep enough to give a good signal-to-noise ratio and enable a uniform velocity profile of the flowing reaction. Depending on the specific mixer design, one might even need channels of varying depth.

Moreover, the selected materials can cause a high scattering background and the production process can induce changes in the material, leading to increased or even non-uniform scattering. This would reduce the overall quality of the data collected in experiments. It is thus of great importance to select a material and production process, which does not lead to a strong scattering background, but still allows producing the required channel geometry.

Here we present the results of our investigations of two production methods that we found suitable for creating microfluidic devices applicable in SAXS experiments: laser ablation of polycarbonate (PC) [5] and injection moulding of Topas brand cyclic olefin copolymers, specifically TOPAS 5013. Laser ablation was selected as it provides great freedom in designing the features of the device and allows for easy and fast design changes, while injection moulding is an effective method for large scale fabrication of Lab-on-chip systems [6,7]. The latter method is less flexible than laser ablation as a stamp must be produced for any given design, but then allows easier and faster production of the microfluidic devices.

# 2 Materials and Methods

## 2.1 Laser ablation

The channels for the microfluidic system were carved into polycarbonate (GoodFellow, UK) by sublimating the material with a Pharos Yb:KGW solid state laser (Light Conversion Ltd., Lithuania), with a wavelength of 1026 nm, a pulse length of 170 fs, at a power of 20 mW and a repetition rate of 200 KHz. The scanning speed was 40 mm/s and the laser spot size 1 µm, being precise enough to resolve the mixer geometry, which is of the order of hundreds of micro-metres. Due to the 3D-sheathing, inaccuracies or surface roughness should have no influence on the mixing or the reaction.

We tested two 1 mm thick plates of PC, one of which was heat treated after the ablation process for 40 minutes at 150°C, which is slightly above the glass transition temperature, in order to reduce surface roughness. Both had five laser ablated square spots with a size of 1x1 mm², which were each ablated a different number of times (from one ablation for the first spot till five ablations for the fifth spot), each with 1000 lines per square (see Fig. 1b). The arrow with the acronym LA indicates the direction of those lines. This resulted in channels of different depth, allowing us to investigate the effect of the laser ablation on the structure of the channel and the resulting impact on the scattering background.

2.2 Injection Moulding

Injection moulding of TOPAS 5013 was done with a Victory 80/45 Tech injection moulding machine (ENGEL, Austria). The initial temperature in the cavity was 280°C and the mould temperature was 130°C (corresponding to the glass transition temperature). The stamp used for the moulding was micro-milled from aluminium. The investigated sample was a 25x75 mm² Topas slide of 1 mm thickness containing a channel shaped like a ring with a gap, with an outer diameter of 10 mm, a channel width of 0.6 mm and a channel depth of 0.1 mm (see Fig. 1c). The direction of the injection moulding is indicated by the arrow with the acronym IM. In order to limit the scanning range we constrained the investigation to the half of the ring containing the ends of the channel.

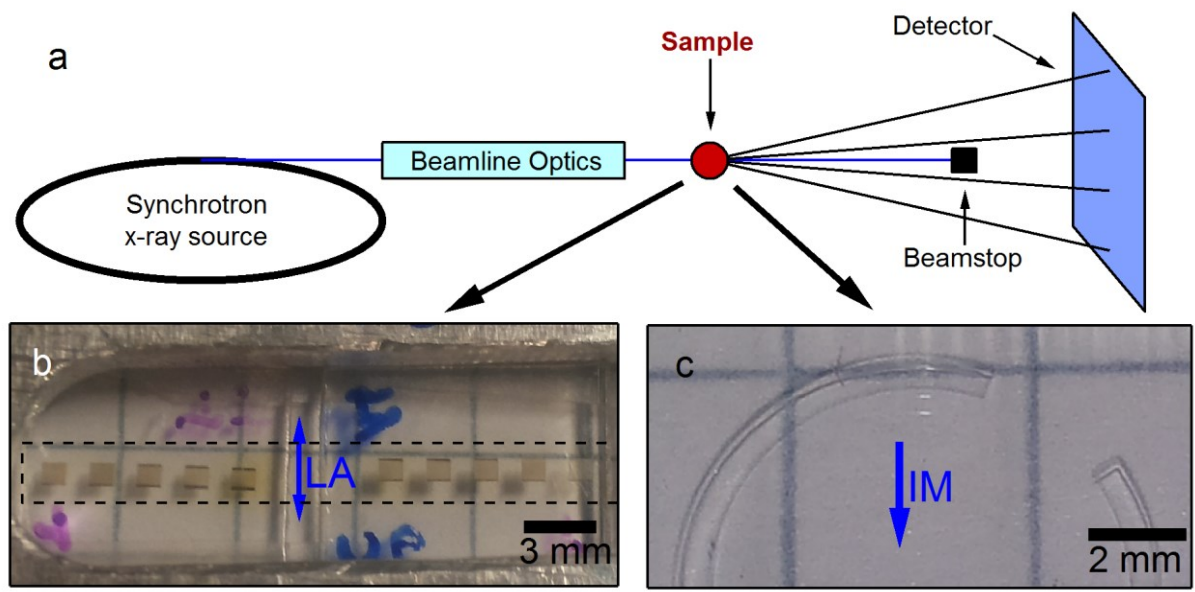

Figure 1: (a) Sketch of the experimental setup. (b) Image of the PC samples with the laser ablated spots as positioned for the measurement (scanning area indicated by the dashed rectangle). The heat-treated sample is on the left and the untreated one on the right. The number of stacked laser passes (direction indicated by LA) increases towards the middle of the two ablated samples. (c) Image of the ring shaped channel in Topas, injection moulded in the direction indicated by IM. The image shows the entire scanning area.

2.3 SAXS measurements

The measurements were carried out at the Austrian SAXS beamline [8] at the Elettra-Sincrotrone Trieste, Italy, with a photon energy of 8keV (sketch shown in Fig. 1a). The sample was mounted on an x-y motor stage, having a positioning precision of 5 μm. The setup allowed for a scan of the entire ablated area. The step size was 250 μm in both directions, with an x-ray beam size of about 300 × 300 μm² (determined with a knife edge scan). The selected scan parameters ensure an overlap of neighbouring scan steps and therefore prevent missing features, which might be critical for the analysis. Additionally, the two spots, which were ablated five times, were also scanned with a smaller step size of 100µm in each direction to achieve a higher resolution.

Scattering patterns were collected with a Pilatus3 1M detector (Dectris, Switzerland), set up at a distance of 1.262 m. The 2D images were averaged using the program Fit2D [9], converting them into 1D scattering patterns $I(q)$, i.e. the scattering intensity as a function of the magnitude of the scattering vector $\vec{q}$,

$$q = \frac{4\pi \sin \theta}{\lambda},$$

where $2\theta$ is the scattering angle and $\lambda$ the wavelength of the x-rays. The effective q-range of 0.1 to 4 nm$^{-1}$ was calibrated using silver behenate [10]. All scattering curves were normalized and the background, obtained from the scattering pattern of the untreated material, was subtracted. With the background subtraction the scattering of the base material as well as the air scattering were removed. The remaining scattering signal is directly due to the changes in scattering intensity introduced by the production process. The plots of the scans in the following discussion show the integrated intensity $\bar{I}$ of the scattering at the respective scan positions [11],

$$\bar{I} = \int I_b(q) q \, dq,$$

with $I_b$ being the normalized, background subtracted scattering intensity, being integrated over the whole q-range of 0.1 to 4 nm$^{-1}$. The integrated intensity plot gives an intuitive overview as a measure of the scattering at a given position.

The measured scattering was absolutely calibrated, to get the differential scattering cross section $d\Sigma(q)/d\Omega$ per unit volume, which is the number of x-rays scattered to a unit solid angle per unit time, unit volume and unit incident flux. The absolute calibration was done using glassy carbon as a secondary standard, which itself was calibrated as described in the supporting information.

From this the differential scattering probability $S(q)$ was subsequently calculated [12],

$$S(q) = \sum_{i \in M} D_i \left(\frac{d\Sigma}{d\Omega}\right)_i (q),$$

where $D_i$ is the thickness of the material $i$. The sum runs over all the consecutive parts that constitute the micromixer along the x-ray path, i.e. channel walls, sample, buffer and sheathing fluid. The patterns in Fig. 2 and Fig. 5 are essentially the probability for photons to be scattered to a given angle, i.e. $q$.

2.4 Depth profiling

Both the laser ablated spots and the injection moulded ring were investigated with an Alpha-Step 500 surface profiler (KLA-Tencor, USA) in order to determine the depth and and roughness of the channels.

The laser ablated channels were probed with the profiler stylus over a range of 1 mm. The scans were started 300 µm outside the channel to enable levelling of the profile. Such profiles were taken in the direction of the laser ablation and perpendicular to it, at a quarter, half and three quarters of the respective channel length.

For the depth, the height of the levelled profile outside the channel was compared with the average height inside, determined visually. Finally the average as well as the standard deviation were calculated from the resulting six values of the depth measurements. For the roughness the difference between the highest peak and the lowest valley was determined inside the channel in each profile and again the average and the standard deviation were calculated.

The injection moulded channel was probed at four positions along the ring with 90° offset to each other. At each of the four positions profiles perpendicular to and along the channel were taken over the length of 1 mm. Depth and roughness were determined as for the laser ablated channels.

## 3 Results and Discussion

3.1 Laser ablation

As observed in Fig. 1b the PC appears less transparent to visible light after the ablation process. As seen in Fig. 2, in which the scan results are presented, this is also true for x-rays. The scattering of x-rays is increased at first, evident in particular in the one-pass spots in Fig 2a. At 1 mm thickness the PC itself is responsible for significant scattering, which is reduced by the removal of material by the ablation, as the scattering is proportional to the thickness [1]. Thus the overall scattering background is reduced despite the increase of the scattering coming from the ablation induced surface roughness (see Fig.3a).
Notably the strong scattering occurs mainly at the channel walls, especially those perpendicular to the writing direction (LA) and is far less pronounced in the centre of the channel, where the walls are not hit. This is ideally the case for any measurement carried out with a carefully designed device, i.e. avoiding the strong scattering of the walls impacting the signal-to-noise ratio.

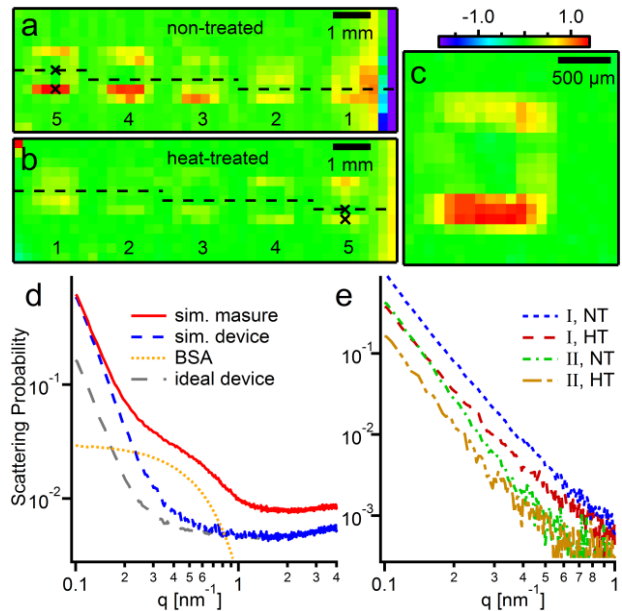

Figure 2: Image of the integrated intensity of (a) the non-treated sample, (b) the heat-treated sample and (c) a high resolution scan (step size of 100 μm) of the non-treated, five times ablated spot. The scattering signal to the far right of (a) and (b) is due to the edge of the PC plates being located there. The dashed lines indicate the position of the horizontal cut shown in Fig. 3a. (d) Scattering patterns of a virtual measurement of BSA (5mg/ml) with a simulated device, as described in the text. It demonstrates the process induced increased scattering, which is notably higher than in an ideal instrument without those effects. (e) The scattering signal at the border (I) and in the centre (II) of the five times ablated spots, indicated by crosses in (a) and (b), for the non-treated (NT) and the heat-treated (HT) samples.

The scattering differs significantly for walls perpendicular to each other. The horizontal walls (as seen in Fig 2) are those where the moving laser changes direction when going from one side to the other. There the walls are less accurately ablated and consequently induce stronger scattering. The vertical walls on the other hand are more precisely cut and have a far weaker influence (compare also Fig. 3). Thus with channels produced by passing the laser along the channel, accidentally hitting the walls with the x-ray beam during measurement is far less problematic.

In order to validate the performance, scattering patterns of the final micromixer were simulated with a standard protein solution as sample (Fig. 2d). The measurement channel was assumed to be 1.8 mm deep, with 200 µm thick PC windows on both sides, for which the data on absolute scale collected in our actual experiments were used (giving 'ideal device' in Fig. 2d). One side was taken to be sealed by a PC sheet, the other, which includes the channel, as having been manufactured by laser ablation without heat-treatment. The additional scattering induced by the ablation increases the background compared to the material in the low-q region by a factor of about 3.5 (at 0.1 nm$^{-1}$).

To investigate the viability of the final micromixer a 5 mg/ml bovine serum albumin (BSA) protein measurement was simulated in the fictive device. The BSA pattern was taken from the SASBDB [13] (SASDA32 at a concentration of 5 mg/ml in 50 mM HEPES 50 mM KCl, pH 7.5) and scaled to absolute scattering units [14]. The measurement assumed 1.2 mm of BSA sheathed on top and bottom by 0.3 mm of water each (the 3D-sheathing). The water was taken to have a constant scattering of 0.01632 cm$^{-1}$ [15], while additional air scattering was neglected. The simulation of the differential scattering probability [12] demonstrates it is well above the background in the Guinier regime of the BSA scattering used for the determination of the radius of gyration.

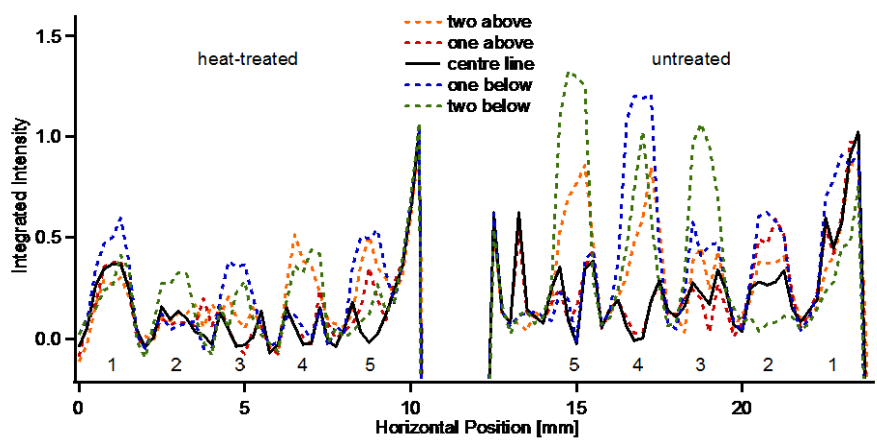

Figure 3: Horizontal cut of the integrated intensity through the middle of all spots (as indicated in Fig. 2). The off-centre scattering (dashed lines), which is due to the horizontal walls is increasing with the number of ablation steps. Meanwhile the centre (thick line) and the vertical walls (peaks of the thick line around the spots) display relatively low scattering. The peaks at the ends are due to the scattering induced by the edges of the PC plates.

Thermal annealing after the laser ablation process reduces the background scattering and thus seems to be recommended. However, the surface characterisation shows that the roughness of the channel floor is higher in the heat-treated sample, increasing up to a value of 1.2 µm. This means that the reduced scattering of the heat-treated sample is due to the relaxation of internal stresses induced by the ablation process.

Unfortunately, heat-treatment also limits the channel depth (see Fig. 4a). While in the untreated case the depth increases more or less linearly with the number of ablation steps, the maximum depth reached by the spots on the heat-treated sample is around 20µm. This means that heat-treatment is not viable for improving micro-channels produced via laser ablation.

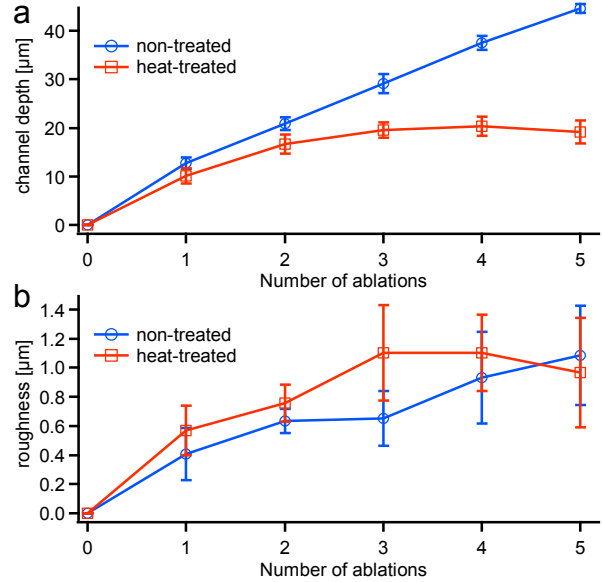

Figure 4: (a) Depth and (b) surface roughness and their standard deviation (indicated by bars) of the laser ablated spots against the number of ablation steps. The depth increases linearly without heat-treatment, but is severely limited after the heat-treatment.

The profile scans indicate an increased ablation in the areas where the laser changes direction. This could explain the increased scattering in these areas. The conical shape of the stylus of the profiler, however, prevents a quantitative analysis of this effect. The increased ablation is present in both the heat-treated and non-treated channels and heat-treatment does not appear to have an effect on it.

3.2 Injection Moulding

We confined the SAXS scanning area to the half of the ring containing the gap (Fig. 1c), with the results shown in Fig. 5. Technical problems at the storage ring led to fluctuations of the intensity during a part of the scan (right side in Fig. 5a). This part, however, still follows the same trends.
Looking at the integrated intensity (Fig. 5a), two effects are immediately apparent: Firstly, there is a lower level of scattering in the channels, as there the amount of material is reduced. Secondly, the amount of scattering varies strongly between different sections of the ring. The horizontal, upper channel wall (A in Fig.5) scatters significantly stronger than the material background, while the vertical walls scatter at the same level as the background.
The background subtracted scattering patterns (see Fig. 5c) reveal that the production process causes no significant additional scattering, when the x-rays hit in the channel centre. A similar virtual measurement as described previously has been simulated and the results are shown in Fig. 5b. With the low material thickness, the process induced scattering becomes notable, but far less than in the case of laser ablation. The device is thus better suited for measuring low scattering samples.

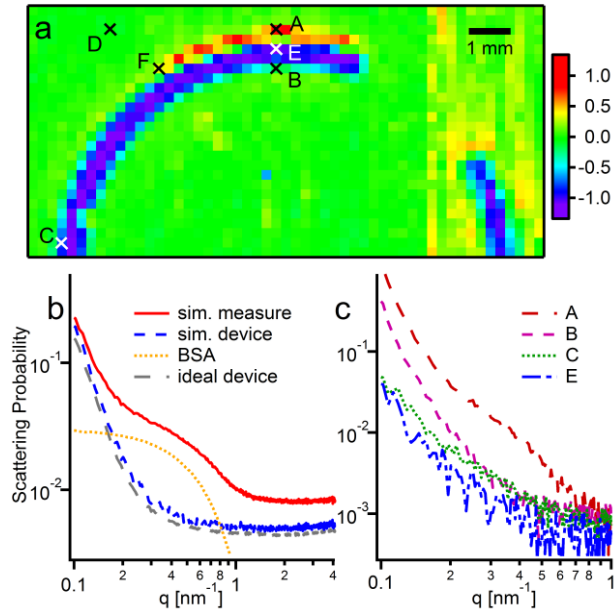

Figure 5: (a) Integrated intensity of the scan over the injection moulded ring. The area on the right side is affected by storage ring based intensity fluctuations, which occurred during the scan. The crosses indicate the selected positions of the patterns that are shown in (c) as well as the scattering images that are shown in Fig. 6. (b) Scattering patterns of a virtual measurement of BSA (5mg/ml) with a simulated device, as described in the text. The process induced scattering is of the same order as the scattering of the material itself. (c) Background subtracted scattering patterns of selected positions, marked in (a). The horizontal walls (A, B) give a far stronger scattering signal than the vertical walls (C) and the channel centre (E).

Similar as in the laser ablated sample, the horizontal walls display stronger scattering than the vertical walls. Examining the raw scattering images, as shown in Fig. 6, reveals that this is due to streaks in the low-q region. The streaks are indicative of reflection and diffraction of the x-rays at the channel walls, as they are always perpendicular to the respective wall.

For vertical walls the streak is covered by the beam stop (Fig. 6d) and thus vanishes from the integration area. This means that the increased scattering is not due to the production process as in the case of laser ablation, but to the orientation of the channel walls with respect to the experimental set-up.

A measurement of the roughness of the channel floor shows fluctuations of the channel depth of up to 50 nm, significantly better than the values achievable by the laser ablation.

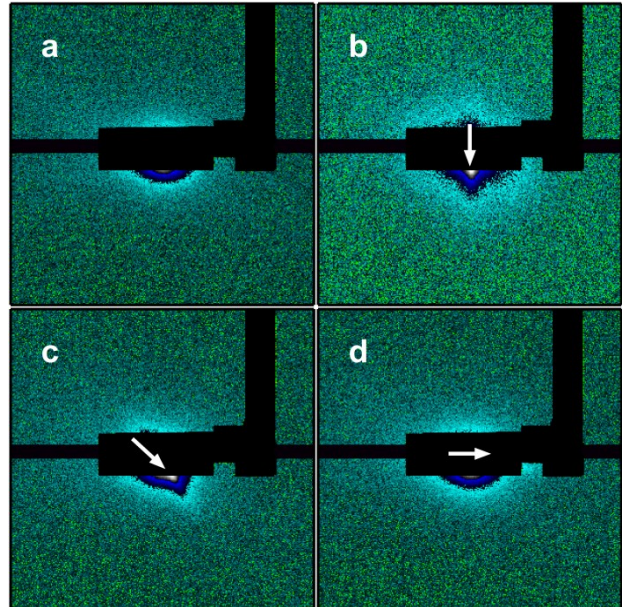

Figure 6: Central region of the scattering images of several positions (q-range from -1 to +1 nm$^{-1}$ horizontally and vertically). (a) Inside the channel, (b) horizontal wall, (c) diagonal wall, (d) vertical wall, all positions marked by crosses in Fig. 4a (E, A, F, C, respectively). The black regions are the masked beam stop and a gap between the detector modules. The streak perpendicular to the direction of the respective wall is indicated by the arrows. For the vertical wall it vanishes behind the beam stop.

## 4 Conclusions

We present two fabrication processes that we investigated and found suitable for creating (micro-) channels specifically intended for the use in SAXS experiments. When producing a microfluidic device by laser ablation, paying attention to the orientation of the production process of the measurement channel can make the difference between highly scattering and nearly invisible channel walls. Thus the ablation should be done in the direction of the measurement channel. The ablation process induces significant background scattering that cannot be reduced by thermal treatment, as this drastically changes the depth of the channels
In the case of devices produced via injection moulding reflective streaks due to the walls occur and potentially mask the data. However, this can be overcome by designing the experiment properly. I.e. mounting the device with the measurement channel vertically allows integrating over the whole vertical length of the detector image without problem, as streaks caused by the channel walls will be in the horizontal direction.

Both methods give usable channels regardless of the wall production orientation, if hit by the x-ray beam precisely in the centre, but correct alignment minimises the influence of accidentally hitting the channel walls.

While both methods where found suitable, the additional scattering induced by the laser ablation restricts its usefulness. The injection moulding causes no significant increase in scattering, thus making it the preferable method. Further improvements could be introduced by decreasing the roughness of the stamps used for injection moulding. This could be achieved by means of stamp-fabrication techniques like dry etching [6,7,16] or Deep X-Ray Lithography (DXRL) [17].

**Acknowledgments**

The authors would like to acknowledge Christian Morello for his technical support throughout the project, as well as Barbara Sartori for useful discussion.
Funding: This project was financed by the H2020-INFRAIA project NFFA-Europe [grant number 654360].